# A 120 lines code for isogeometric topology optimization and its extension to 3D in MATLAB

Xianda Xie[a], Zhihui Ou[a], Aodi Yang[c], Xiaobing Li[a], Shuting Wang[b],*

[a] *School of Advanced Manufacturing，Nanchang University, Nanchang, 330031, China*
[b] *School of Mechanical Science and Engineering, Huazhong University of Science and Technology, Wuhan, 430074, China*
[c] *Shenzhen Xinkailai Technology Co Ltd, Shenzhen, 518111, China*

**Abstract**

In this paper, a compact and efficient code implementation is presented for isogeometric topology optimization (ITO) approach. With the aid of Bėzier extraction technique, a derived explicit stiffness matrix computation formula is applied to all B-spline IGA elements with rectangular shape under linear elasticity assumption. Using the aforementioned explicit formula, the stiffness matrix calculation and updating of IGA are significantly simplified, which leads to the current ITO code implemented only in one main function without calling subroutines, such as IGA mesh generation and Gaussian quadrature. Both two-dimensional (2D) and three-dimensional (3D) cases are taken into consideration, which result into iga_top120 and iga_top3D257 MATLAB codes for 2D and 3D design problems. Numerical examples validate the effectiveness of our open-source codes, with several user-defined input parameters basically identical to those used in top88 and top3D. Therefore, iga_top120 and iga_top3D257 provide an effective entry for the code transforming from FEM-based TO into ITO.

## 1 Introduction

Topology optimization (TO) launched by Bendsoe and Kikuchi [1] has achieved extensive improvements in the past three decades, which gains the popularity from academic to engineering fields due to its effectiveness in designing various engineering structures. Several canonical TO approaches have been put forward with different descriptions of design variables, such as Solid Isotropic Material with Penalization (SIMP) [2, 3], Evolutionary Structural Optimization (ESO) [4], Level Set Method (LSM) [5, 6], Moving Morphable Components/Voids (MMCs/MMVs) [7, 8]. Following the noble way of sharing 99-line MATLAB code for SIMP method [9], the educational MATLAB codes of various TO methods have been published over past decade, such the improved MATLAB codes for SIMP method [10-12], a MATLAB code for bidirectional ESO [13], the 88-line MATLAB code for parameterized LSM [14], and the MATLAB code for explicit TO using MMCs [15]. Besides, the open-source codes for topology optimization considering multi-material design [16], stress constraint [17], geometric nonlinearity [18], and etc., are summarized in [19].

In these codes mentioned above, traditional Finite Element Method (FEM) with Lagrange basis functions plays the role of solver for the structural response of TO in an iterative manner, and the low order elements are adverse to the numerical stability of the overall optimization process, while the high order elements have side effect on solution accuracy [20]. As an advanced FEM technique, Isogeometric Analysis (IGA) initialized by Hughes et al. [21] opened the door to the seamless integration between Computer Aided Design (CAD) and Computer Aided Engineering (CAE), which improves the numerical accuracy of FEM and the computational efficiency for high order elements. To integrate with current mature FEM commercial software, finite element data structure was developed for NURBS-based IGA [22] through Bézier extraction technique, and its extension to T-spline-based IGA was elaborated in detail in [23]. Therefore, the application of IGA technique in topology optimization has attracted wide attention over the last decade, due to its demonstrated advantages regarding integration with CAD.

---

* Corresponding author.
  *E-mail address:* wangst@hust.edu.cn (S. Wang).

The isogeometric topology optimization (ITO) can be tracked back into 2010 [24], where the trimmed surface analysis is employed to design domains described by trimmed geometries and the optimized structures can be converted into CAD models effortlessly. Subsequently, Dedè et al. [25] presented a phase model-based TO method with the optimal topology obtained by the steady state of the phase transition, and constructed the spatial approximation by IGA suitable for solving phase field problems particularly. Qian [26] proposed an implicit technique for ITO by exploiting the local support property of B-spline, which has been significantly improved by the tensor product splitting technique in a recent work [27]. Wang et al. [28] make a comparatively complete comparisons between IGA and conventional FEM method within a parametric LSM TO framework, which advocate the superiority of IGA over FEM in high order elements. Hou et al. [29] put forward explicit ITO method using MMCs description, of which the continuity of topology description functions was promoted by R-functions within the overlapped regions between different MMCs [30]. Zhang et al. [31] presented an explicit ITO framework using MMV topological description for shell structures design problems. Yang et al. [32] devised a post-processing technique for ITO approach, where the editable CAD models of optimal designs can be automatically constructed. To deal with design problems with complex design domain, Gao et al. [33] proposed an ITO method in terms of NURBS-based finite cell method, of which the essential boundary condition is tackled by Nitsche's method [34] in a variationally weak sense. More recently, the complex shell structure design problems are optimized by ITO using NURBS-based finite cell method in [35], and a flexible ITO design approaches is developed for cellular structures by the multi-patch technique [36]. After nearly ten years of development, the ITO method has been applied to practical structural design problems, such as material design problems [37-40], stress-related design problems [41, 42], geometrically nonlinear problems [43], and structural dynamic design problems [44].

However, the deficiency of ITO in efficiency is intrinsically inherited from the use of high order elements and tensor product structure of NURBS. Wang et al. [45] proposed an efficient ITO method with the use of multilevel design mesh, and multigrid conjugate gradients solver, as well as local update strategy designed for optimization variables. Xie et al. [46] put forward an adaptive ITO approach in terms of truncated hierarchical B-splines, which strikes a good balance between optimization accuracy and solution efficiency. Gupta et al. [47] applied the PHT spline technique to ITO, which achieves adaptive discretization and significant reduction in CPU time than the non-adaptive case. Yang et al. [48] derived an explicit stiffness matrix expression of IGA element for ITO, which can simplify the preprocessing procedure of ITO to a large extend when uniform regular IGA mesh is used. More recently, an object-oriented ITO software is presented in [49], which can extend ITO into stiffness design problems with complex design domain by means of immersed boundary method. Although the open-source codes for IGA have been presented in [50-52] and the MATLAB codes for ITO have been reported in [53-55], there is still lacking a compact and efficient implementation for ITO even for rectangular IGA mesh and it hinders newcomers to understand the basic fundamentals of ITO [19].

The main contribution of this work lies in a 120 lines MATLAB code for ITO solving two-dimensional (2D) compliance minimization problem, and the associated code for three-dimensional (3D) classical TO problems, which can be accessed by https://github.com/NcuMikeXie/igatop conveniently.

These two codes are the implementations of the ITO method proposed in [48], using the explicit stiffness computation for design domains with rectangular or cuboid shapes. In the presented MATLAB codes, the troublesome subroutines are avoided in comparison with education ITO codes in [53], such as the derivation of B-spline basis function and assembling global stiffness matrix. Therefore, this work bridges FEM-based TO educational codes to ITO framework with minor modifications. The remaining content of this paper is organized as follows: the key ingredients of ITO using Bėzier extraction operator are briefly introduced in Section 2, and Section 3 makes a detailed explanation for the 120 lines MATLAB ITO code. Several classical TO benchmarks are presented in Section 4 to illustrate the effectiveness of presented ITO codes, and the conclusions of this work are drawn in Section 5.

## 2 Problem formulation of isogeometric topology optimization

In this section, we discuss the theoretical aspects of ITO, which lay a solid foundation for the presented 120 lines code. In particular, we present Bėzier extraction operator for B-splines, the idea of the explicit stiffness matrix computation for rectangular B-spline IGA elements, as well as the ITO model for 2D compliance problems. For the state of art of this method, interesting readers can refer to [48] for further details.

### 2.1 Bėzier extraction operator

Given an open knot vector $\Xi = \{\xi_0, \cdots, \xi_{n+p+1}\}$ with non-decreasing sequence of real numbers, a B-spline basis function $b_{i,p}(\xi)$ can be constructed from basis functions with $p = 0$ using the classical Cox-de Boor recursion formula, as follows:

$$b_{i,0}(\xi) = \begin{cases} 1 & \xi_i \leq \xi < \xi_{i+1}, \\ 0 & \text{otherwise}, \end{cases}$$
$$b_{i,p}(\xi) = \frac{\xi - \xi_i}{\xi_{i+p} - \xi_i} b_{i,p-1}(\xi) + \frac{\xi_{i+p+1} - \xi}{\xi_{i+p+1} - \xi_{i+1}} b_{i+1,p-1}(\xi). \tag{1}$$

Then, a spline curve $\tau$ can be defined by a set of control points $P_i$ with $\tau(\xi) = \sum_{i=1}^{n} P_i \cdot b_i(\xi)$, of which the Bézier decomposition is generated by raising the multiplicities of all internal knots from 1 to $p$. As a consequence, the B-spline curve $\tau$ is equivalently represented by several piecewise Bézier curves in terms of Bernstein polynomials $B_i(\xi)$, as $\tau(\xi) = \sum_{i=1}^{\bar{n}} \bar{P}_i \cdot B_i(\xi)$. A linear transformation exists between the original B-spline control points and the Bézier control points [22], which is formulated as a matrix operation:

$$\bar{P} = C^T P. \tag{2}$$

In Eq. (2), $P = \{P_i\}$ and $\bar{P} = \{\bar{P}_i\}$ are the union of B-spline and Bėzier control points, as well as $C^T$ is termed as Bézier extraction operator, of which the components is obtained from the Boehm's algorithm [56]. Substituting Eq. (2) into $\tau(\xi) = b^T P = B^T \bar{P}$ with $b = \{b_i(\xi)\}$ and $\bar{b} = \{\bar{b}_i(\xi)\}$, we can obtain the following equation:

$$\tau(\xi) = b^T P = B^T \bar{P} = B^T (C^T P) = (CB)^T P. \tag{3}$$

The arbitrariness of $P$ yields:

$$b = CB. \tag{4}$$

Therefore, the B-splines basis functions can be related with the Bernstein basis functions through the associated decomposition, which can be illustrated in Fig. 1.

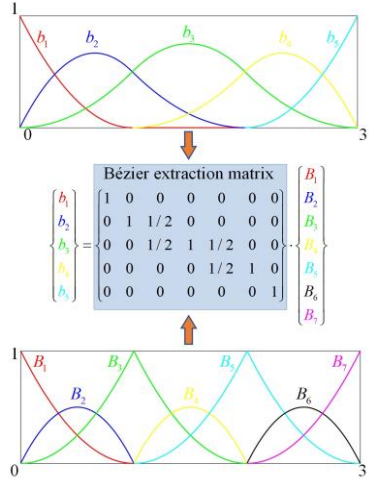

**Fig. 1** Illustration of Bézier extraction operator for B-spline basis functions of the open knot vector $\varXi = \{0,0,0,1,2,3,3,3\}$.

**2.2 Explicit stiffness matrix computation for B-spline elements**

To facilitate the generation of stiffness matrices of B-spline elements, an explicit computation formula is derived for rectangular B-spline elements, in terms of the Bézier extraction operators along the parametric directions of IGA mesh and the standard Bézier stiffness matrix, rather than implicitly obtained by performing Gaussian quadrature process on each B-spline element. This explicit formula has been derived exhaustively in [48] and formulated as:

$$\mathbf{K}_e = \boldsymbol{C}_e \cdot \mathbf{K}_{Bézier}^0 \boldsymbol{C}_e^{\mathrm{T}}, \tag{5}$$

where the stiffness matrices of the *e*-th B-spline element and standard Bézier element are represented by $\mathbf{K}_e$ and $\mathbf{K}_{Bézier}^0$, respectively. $\boldsymbol{C}_e$ is termed as the transformation matrix between these two stiffness matrices, which is calculated as the Kronecker product of the elemental univariate Bézier extraction matrices $\boldsymbol{C}_{e,\eta}$ and $\boldsymbol{C}_{e,\xi}$, and written in:

$$\boldsymbol{C}_e = \begin{bmatrix} \boldsymbol{C}_{e,\eta} \otimes \boldsymbol{C}_{e,\xi} & \mathbf{0} \\ \mathbf{0} & \boldsymbol{C}_{e,\eta} \otimes \boldsymbol{C}_{e,\xi} \end{bmatrix}. \tag{6}$$

**2.3 Isogeometric topology optimization for compliance problem**

Given a design domain subjected to an external load, the compliance problem of ITO is defined as seeking the optimal material distribution with the structural compliance minimized and a volume constraint applied to the solid material usage. Therefore, the objective of ITO model is the strain energy of the structure, which is equivalent to the external work of external load and presented in the form of:

$$c(\bar{\boldsymbol{x}}) = \boldsymbol{f}^{\mathrm{T}} \cdot \boldsymbol{u}(\bar{\boldsymbol{x}}). \tag{7}$$

In Eq. (7), *c* represents the structural compliance, *f* is termed as the external load vector, *u* denotes the displacement vector of all degree of freedoms (DOFs). Then, the mathematical model of ITO is formulated as:

$$\begin{aligned}
&\text{find } \bar{x} = (x_1, x_2, x_3, \ldots, x_{nelx \cdot nely}), \\
&\text{minimize } c(\bar{x}) = \boldsymbol{f}^{\text{T}} \cdot \boldsymbol{u}(\bar{x}) = \sum_{j=1}^{nely} \sum_{i=1}^{nelx} E_{i,j}(x_{i,j}) \cdot (\boldsymbol{C}_{i,j}^{\text{T}} \boldsymbol{u}_{i,j})^{\text{T}} \cdot \mathbf{K}_{Bézier}^{0} \cdot (\boldsymbol{C}_{i,j}^{\text{T}} \boldsymbol{u}_{i,j}), \\
&s.t. \\
&\mathbf{K}(\bar{x}) \cdot \boldsymbol{u}(\bar{x}) = \left( \sum_{j=1}^{nely} \sum_{i=1}^{nelx} E_{i,j}(x_{i,j}) \cdot \boldsymbol{C}_{i,j} \cdot \mathbf{K}_{Bézier}^{0} \cdot \boldsymbol{C}_{i,j}^{\text{T}} \right) \cdot \boldsymbol{u}(\bar{x}) = \boldsymbol{f}, \\
&\frac{A(\bar{x})}{A_0} \leq frac, \\
&\bar{x} \subset \aleph, \ \aleph = \{\bar{x} \in \mathbb{R}^{nelx \cdot nely}, \mathbf{0} \leq \bar{x} \leq \mathbf{1}\}.
\end{aligned} \qquad (8)$$

In Eq. (8), the number of elements along the parametric axes are denoted by *nelx* and *nely*, $x_{i,j}$ is treated as the element-wise constant density design variable of $(i, j)$-th IGA element, $\boldsymbol{u}_{i,j}$ is the local displacement vector, $\mathbf{K}(\bar{x})$ represents the global stiffness matrix for IGA, $A(\bar{x})$ and $A_0$ are the areas occupied with solid material and the whole design domain, *frac* denotes the prescribed upper limit of solid material, $\aleph$ represents the admissible space where the design variable set $\bar{x}$ belongs to. Moreover, the stiffness transformation matrix is denoted by $\boldsymbol{C}_{i,j}$, and the Young's elastic modulus $E_{i,j}(x_{i,j})$ is obtained by the modified SIMP model for the $(i, j)$-th IGA element, which is defined as:

$$E_{i,j}(x_{i,j}) = E_{min} + (x_{i,j})^{pen} \cdot (E_0 - E_{min}), (x_{i,j} \in [0,1]), \qquad (9)$$

where $E_{min}$ and $E_0$ are the elastic modulus for the empty and solid elements with $0 < E_{min} = E_0$, *pen* represents the penalty factor, which equals to 3 in this work for convenience.

To solve the ITO model presented in Eq. (8), a heuristic updating scheme for design variables is derived from the standard optimality criteria method, which is formulated as:

$$x_e^{new} = \begin{cases} \max(0, x_e - m) & \text{if } x_e B_e^{\eta} \leq \max(0, x_e - m), \\ \min(1, x_e + m) & \text{if } x_e B_e^{\eta} \geq \min(1, x_e + m), \\ x_e B_e^{\eta} & \text{otherwise,} \end{cases} \qquad (10)$$

where $e = (j-1) \cdot nely + i$ denotes the global index of the $(i, j)$-th IGA element, $m$ is termed as the positive move limit of all design variables, $\eta$ represents the numerical damping coefficient with $\eta = 0.5$, $B_e$ plays the role of scaling factor and is determinized by the optimality condition:

$$B_e = \frac{-\frac{\partial \hat{c}}{\partial x_e}}{\lambda \frac{\partial A}{\partial x_e}}. \qquad (11)$$

In Eq. (11), $\lambda$ represents the Lagrangian multiplier for applying the volume constraint of ITO, whose appropriate value is determined by a bisection algorithm to fulfilling the volume constraint.

The filtering sensitivity $\partial \hat{c} / \partial x_e$ of the objective function is given by:

$$\frac{\partial \hat{c}}{\partial x_e} = \frac{1}{\max(\gamma, x_e) \sum_{s \in N_e} w_{es}} \sum_{s \in N_e} w_{es} x_s \frac{\partial c}{\partial x_s}, \qquad (12)$$

with the original sensitivity $\partial c / \partial x_s$ of the objective function defined as:

$$\frac{\partial c}{\partial x_s} = -p \cdot (x_e)^{p-1}(E_0 - E_{min}) \cdot (\boldsymbol{C}_s^{\mathrm{T}}\boldsymbol{u}_s)^{\mathrm{T}} \cdot \boldsymbol{K}_{B\grave{e}zier}^0 \cdot (\boldsymbol{C}_s^{\mathrm{T}}\boldsymbol{u}_s). \tag{13}$$

Moreover, $w_{es}$ is termed as the weight factor of the $s$-th element objective sensitivity contributing to the $e$-th element, which is calculated by:

$$w_{es} = \begin{cases} r_{min} - \mathrm{dist}(e, s), & \text{if } \mathrm{dist}(e, s) \leq r_{min} \\ 0, & \text{otherwise} \end{cases} \quad (e = 1, \ldots, nelx \cdot nely), \tag{14}$$

where $r_{min}$ is a user-defined sensitivity filter radius, $\mathrm{dist}(e,s)$ represents the centroid distance between these two design elements.

Besides, the sensitivity $\partial A / \partial x_e$ of volume constraint equals to 1, if it is supposed that each element has unit area.

## 3 Matlab implementation

In this section, we introduce the 120 lines MATLAB code for ITO (see Appendix A) in detail, of which the overall flowchart is illustrated in Fig. 2. Similar to the classical topology optimization MATLAB codes, the ITO code is invoked by the following MATLAB command line:

iga_top120(nelx, nely, volfrac, penal, rmin)

where nelx and nely are the number of IGA elements along the $\xi$ and $\eta$ parametric directions, respectively, volfrac represents the upper bound of volume fraction for solid material usage, penal denotes the penalty factor $p$ used in Eq. (9), rmin is termed as filter radius $r_{min}$ ( divided by the IGA element size). Differing from the way of using an additional argument to specify the type of filtering technique, we use the sensitivity filter in our current ITO implementation by default.

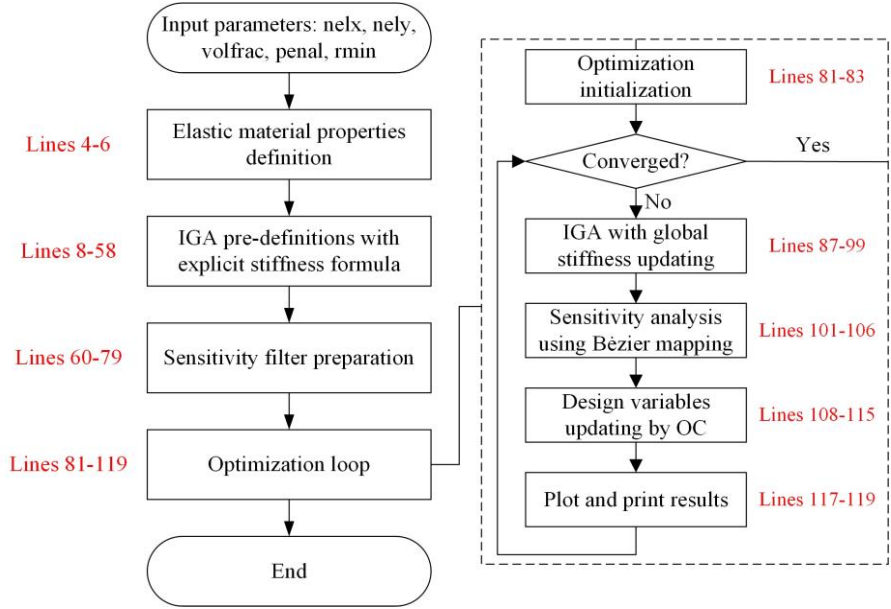

**Fig. 2** Overall flowchart of iga_top120 based on an explicit stiffness matrix formula for B-spline IGA elements.

The most obvious differences between the 120 lines code and the current open-source ITO codes [53, 54] are listed as: 1) the subroutines used to calculate the derivatives of shape functions and the elemental stiffness matrices using Gaussian quadrature, are replaced by several lines code implementing the presented explicit stiffness matrix computation; 2) the element connectivity matrices are efficiently generated by the IGA mesh size and degrees of B-spline, without resorting to the knot interval searching algorithm; 3) similar to most of FEM-based TO codes, the iterative optimization loop is compact and maximum amount is only executed once, which provides a convenient integration of FEM-based TO algorithms into ITO framework; 4) an extra definition for element connectivity matrices of the decomposed Bézier mesh is included in order to facilitate the computation of the structural compliance in a vectorized manner; 5) a main program implements the ITO without additional subroutines; 6) the input parameters are basically identical to those used in top88 [10].

Similar to the widely spread top88 [10], the presented 120 lines ITO code consists of the following three parts: isogeometric analysis, sensitivity filter and optimization loop, which are introduced thoroughly in subsections 3.1-3.3.

### 3.1 Isogeometric analysis (IGA)

For the classical topology optimization problem of MBB beam illustrated in Fig. 3 (a), the design domain is prescribed to be discretized into a rectangular IGA mesh with square B-spline elements, which is treated as a prerequisite, when 120 lines ITO code is used to solve the TO problems. A coarse example IGA mesh consisting of 8 quadratic B-spline elements is shown in Fig. 3 (b), with nine control points determining per element and two degrees of freedom (DOFs) assigned to per control point. Both the control points and elements are numbered in row-wise from left to right, with the row staring from the bottom. Slightly different from the DOFs encoding scheme used in top88 [10], we follow the DOFs encoding scheme used in [52], and the DOFs $n$ and $n + noCtrPts$ correspond to the horizontal and vertical displacement of the *n*-th control point with *noCtrPts* representing the number of control points for IGA mesh. By taking advantage of the high regularity of IGA mesh, the computational effort in the optimization loop can be reduced in several different aspects and achieves a minimum phase.

Following the most of FEM-based TO codes, the isogeometric analysis preprocessing part defines the material physical properties (lines 4-6) firstly: E0 denotes the Young's modulus $E_0$ of solid material, Emin represents the Young's modulus $E_{min}$ of the artificial material for void regions, and nu is treated as the Poisson's ratio *v*. Several parameters related to B-spline and Bézier IGA meshes are computed (lines 8-15), such as the number of B-spline and Bézier control points along the parametric directions as well as the B-spline IGA elements.

Next the stiffness matrix of Bézier element is defined by the Young's modulus equaling to 1 (lines 16-29), which is represented by KE. According to the derivation presented in [48], the stiffness matrices of all Bézier elements can be standardized by KE, when the IGA mesh regularity is guaranteed. To implement the explicit formula presented in Eq.(5), the fundamental matrices for stiffness transformation matrix are required to be pre-defined (lines 30-41), which are the key ingredients for the calculation of B-spline IGA elements in the optimization loop.

In order to allow an efficient computation of the objective function sensitivity shown in Eq.(13), the DOFs index matrix edofMat_B of Bézier control mesh is constructed (lines 42-45). The similar situation occurs for the efficient assembly of the IGA stiffness matrix in the optimization loop, where the DOFs index matrix edofMat of B-spline control mesh is prescribed (lines 46-49). For edofMat, its *i*-th row contains eighteen DOFs indices corresponding to the *i*-th B-spline IGA element. The matrix edofMat is constructed by the following three steps. Firstly, the parametric connectivity matrices elConnU and elConnV are defined by the input parameters nelx and nely, where the MATLAB function bsxfun is used to return the element connectivity matrices with the size determined by the second and third input vector parameters automatically, whose elements are calculated by applying plus operation to the column-constant matrix generated by the second input parameter and the row-constant matrix generated by third input parameter. Secondly, the element connectivity matrix is obtained by performing repeating and permuting and addition operation on the parametric connectivity matrices elConnU and elConnV. Thirdly, the DOFs index matrix edofMat is defined by the element connectivity matrix straightforwardly in line 49. With the aid of edofMat, the horizontal and vertical index vectors ik and jk are calculated on the fly in lines 50-51, which are used to allocate the stiffness values from the local element level to the global level, and facilitate the efficient updating of global stiffness matrix in the iteration loop. Subsequently, the auxiliary variables elConnU_B, elConnV_B, element_B, elConnU, elConnV, element and edofMat are cleared in line 52 for the memory efficiency.

Once the prerequisite of IGA is prepared as mentioned above, the boundary module of IGA is presented in lines 54-58, which is specified to the benchmark of half MBB problem as illustrated in Fig. 3.

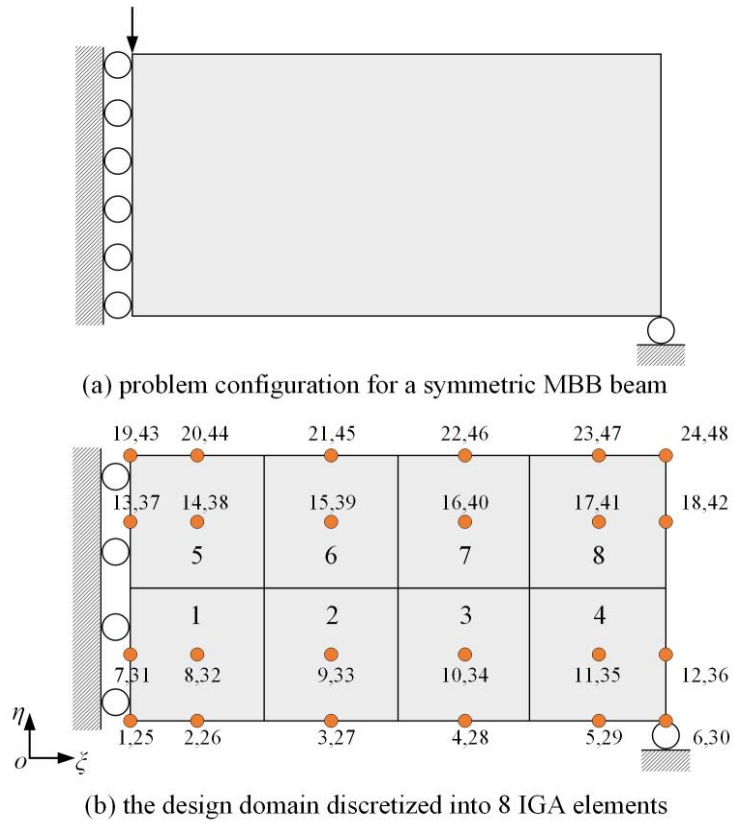

**Fig. 3** Illustration of the classical half MBB beam TO design problem and its coarse quadratic IGA mesh with the size of 4×2.

### 3.2 Sensitivity filter

Either FEM-based or IGA-based TO method, sensitivity filter plays one of the vital roles, as it prevents the optimal design from falling into checkboard pattern and mesh-dependency situations. The lines 60-62 are used to initialize the three basic parameters iH, jH and sH, of which the values are obtained by four loops shown in lines 63-77 with rmin used to define the filter radius. Afterwards, the weigh matrix H is assembled by the sparse command of MATLAB with iH, jH and sH treated as the input parameters in line 78. According to Eq. (12), the modified sensitivity for each design element is calculated as the weighted average of its neighbourhood sensitivities within the filter circle. Therefore, the weight sum vector Hs is obtained by accumulating all column values of the weight matrix H in line 79.

### 3.3 Optimization loop

Ahead of the execution of optimization loop, all design variables are initialized to the prescribed volume fraction limit of the solid material usage in line 81, and the number of loop and the maximum change in design variables between two consecutive iterative steps are set to 0 and 1 in line 82 and line 83, respectively.

The optimization loop starts from the convergence criterion determination with the maximum change in design variables no less than 0.01, as presented in line 84. When the convergence criterion is not satisfied, the program will execute the codes in the optimization loop iteratively (lines 85-120) until the convergence criterion is fulfilled. Basically, the lines 85-120 can be divided into four parts, which are listed as follows: a) isogeometric analysis (lines 86-99); b) sensitivity analysis (lines 100-106); c) updating scheme of design variables (lines 107-115); d) optimization results display (lines 116-119).

In the isogeometric analysis shown in lines 86-99, the distinct feature lies in the explicit matrix computation formula presented in Eq. (5) replacing the traditional tedious full Gaussian quadrature rule, which leads to the compact code implementation illustrated in lines 87-95, and the stiffness values of all IGA elements sK are calculated. By invoking the sparse command of MATLAB, the global stiffness matrix K is assembled in line 96 with iK, jK and sK treated as the input parameters, of which the symmetry is guaranteed by line 97. Then, the displacement vector U of B-spline control points is obtained by applying the "\" operation between global stiffness matrix K and external load vector F, as presented in line 98. Based on Eq. (8), the displacement vector U_B associated with Bézier control points is computed as the linear combination between the Bėzier extraction matrix and U in line 99.

For the purpose of updating the element-wise density design variables, the strain energies ce of all IGA elements are obtained in line 101 with U_B and KE used here, which facilitates the efficient implementation of sensitivity analysis, rather than resorting to the tedious stiffness matrix computations for all B-spline elements. Subsequently, the objective function c is calculated as the sum of all elements strain energies ce of all elements, taking the modified SIMP material model formulated as Eq. (9) into consideration. The sensitivities dc and dv are presented in line 103 and line 104 for the objective function and volume constraint function, respectively. Using the sensitivity filtering technique presented in Eq. (12), the modified sensitivity is obtained in line 106, which is identical to the implementation of top88 [10] for the objective function. The new design variables vector *xnew* is obtained by the optimality criteria method presented in lines 108-113, and the maximum change in design variables is computed in line 114, as well as the design variables vector *x* is assigned to *xnew*.

Finally, the basic metrics of the optimization process are displayed in the MATLAB command window by the code written in line 117, including the loop number, objective function, volume fraction of solid material and maximum change in design variables. The black-and-white optimized structural design is shown in the MATLAB graphical window by line 119.

## 4 Numerical examples

In this section, two benchmarks are presented to demonstrate the effectiveness of the MATLAB codes iga_top120 and iga_top3D257 for two-dimensional (2D) and three-dimensional (3D) B-spline based ITO methods, respectively. Without otherwise specified, the Poisson ratio is set to 0.3, and the Young's moduli are prescribed to 1 and $10^{-3}$ for solid and void material, as well as the penalty factor *pen* used in Eq. (9) takes 3. A laptop with 12th Gen Intel(R) Core (TM) i7-1260P 2.10 GHz and 16GB RAM, is treated as the hardware test environment, and Windows 10 Operating System and MATLAB R2021b construct the software environment. The convergence criterion of the optimization loop is that the $L_\infty$ of the difference between design variable vectors on two consecutive steps is no large than 0.01.

**4.1 Half Messerschmitt–Bolkow–Blohm (MBB) beam**

As illustrated in Fig. 3 (a), the left edge is simply supported in horizontal direction and the right-bottom corner is simply supported in vertical direction, and a downward external load is imposed on left-top corner with the magnitude equaling to 1 for half MBB design problem. The upper limit in solid material volume fraction is 0.5. Four different IGA mesh sizes are taken into consideration here, consisting of 60×20, 150×50, 300×100, and 600×200. Identical to the widespread top88 [10], the filter radius takes 0.04 times the number of elements (or knot spans) in horizontal direction, i.e. 2.4, 6, 12 and 24 for different mesh resolutions.

Fig. 4 shows the optimized topological results of half MBB problem in combination with the corresponding compliance, which are obtained by invoking iga_top120 with five user-defined parameters. These topology results are identical to these obtained by top88 using sensitivity filtering technique illustrated in Figure. 3 in Ref. [10], and the slightly differences in compliance are resulted from the solver type difference. Therefore, iga_top120 can applied to 2D compliance TO problems effectively.

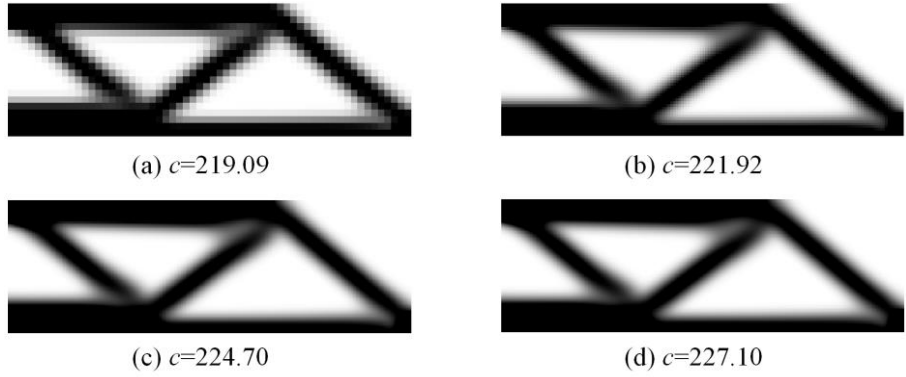

(a) $c$=219.09  (b) $c$=221.92

(c) $c$=224.70  (d) $c$=227.10

**Fig. 4** Illustration of the generated optimal designs by iga_top120 code for half MBB problem: (a) for 60×20 uniform B-spline IGA mesh; (b) for 150×50 uniform B-spline IGA mesh; (c) for 300×100 uniform B-spline IGA mesh; (d) for 600×200 uniform B-spline IGA mesh.

**4.2 Three-dimensional cantilever**

A 3D cantilever is taken into consideration in this subsection to verify the effectiveness of the presented iga_top3D257, of which the problem configuration is illustrated in Fig. 5. The cantilever is fixed on the left shadow face, and an external point load downwardly along Y direction is imposed on the central point of the right face, as well as the upper limit in solid material usage is specified to 0.2. To obtain the displacement field by isogeometric analysis, the cantilever is discretized into uniform B-spline IGA mesh with the size of 40×20×20, and the basis degrees combinations of all IGA elements are set to 2×2×2.

When the sensitivity filter is set to $2\sqrt{3}$, the 3D cantilever TO design problem can be solved by invoking iga_top3D257(40, 20, 20, 0.2, 3, 2*sqrt(3)). Fig. 6 presents the optimized structural topology by iga_top3D257. Moreover, the FEM-based TO method is also applied to optimize the provided design problem, by invoking top3D(40, 20, 20, 0.2, 3, 2*sqrt(3)) [11], where the elastic modulus of the void material $E_{min}$ is changed from $10^{-9}$ into $10^{-3}$ with slightly improved convergence from our numerical experience. The generated structural topology by top3D is presented in Fig. 7, which is similar to the one shown in Fig. 6. According to the results presented in Fig. 6 and Fig. 7, we can find that

iga_top3D257 outperforms top3D in terms of the objective function and the numerical efficiency, as the structural compliance *c* (measured by FEM) and the overall solution time T are decreased by 8.6% and 34.4%, respectively. Fig. 8 makes a comparison between our presented iga_top2D257 and top3D, which indicates a significant improvement in numerical convergence when FEM solver is replaced by IGA for 3D problems. The per iteration solution efficiency of iga_top2D257 is decreased by 81.85% than top3D, which is resulted from the fact that tri-quadratic basis functions are applied in iga_top2D257, while tri-linear basis functions are used in top3D. Finally, iga_top3D257 can be effectively used to solve 3D minimization compliance TO design problems, and achieves a better performance than the FEM-based TO method, which is attributed to the excellent numerical performance of IGA in solution accuracy than FEM with linear interpolation basis.

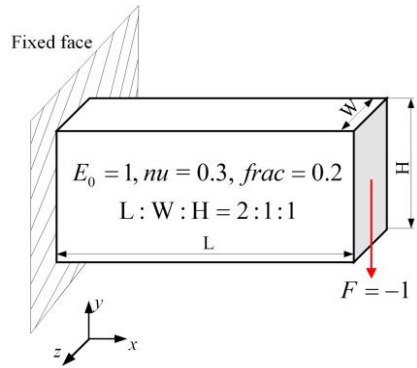

**Fig. 5** Illustration of the problem setting for 3D cantilever subjected to a central point load on right face.

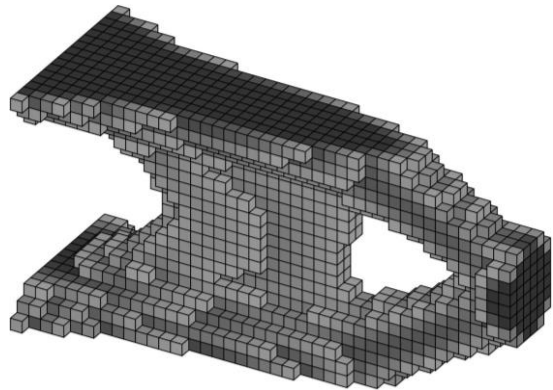

**Fig. 6** Converged ITO design result for 3D cantilever by our presented iga_top3D257 code, with *c*=9.35 (IGA) or *c*=12.62 (FEM) and Iterations=22 as well as the overall solution time T=397.2 (seconds).

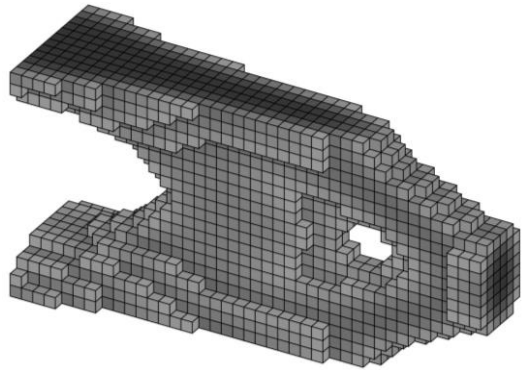

**Fig. 7** Converged FEM-based TO design result for 3D cantilever by top3D code presented in [11], with *c*=13.81 (FEM) and Iterations=184 as well as the overall solution time T=605.6 (seconds).

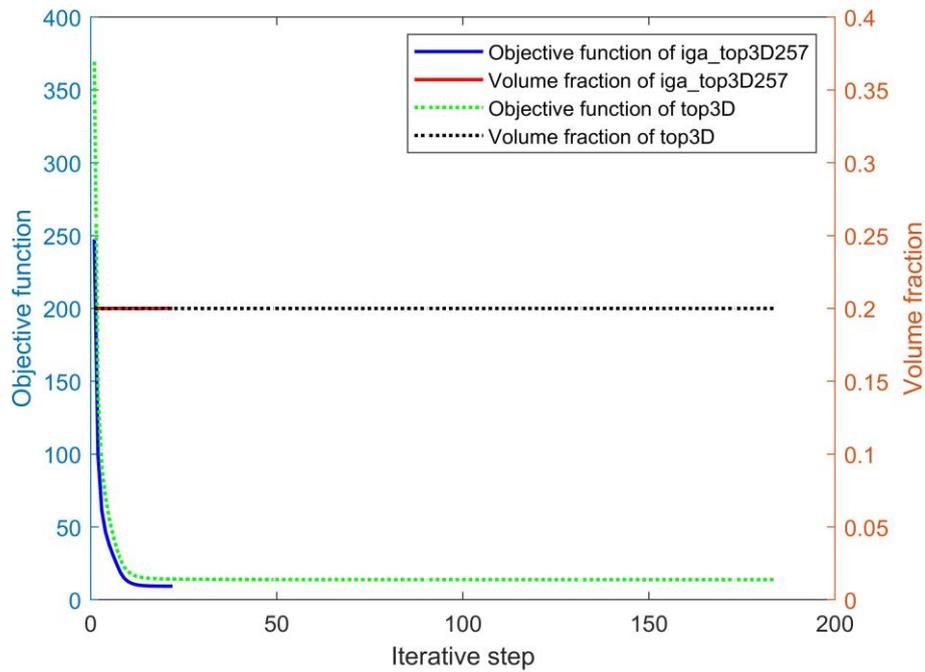

**Fig. 8** Illustration of the comparison in convergence history between iga_top3D257 and top3D for the 3D cantilever TO design problem.

## 5 Conclusions

This work presents two open-source codes in a main function for isogeometric topology optimization (ITO) written in MATLAB for 2D and 3D TO problems, which have merits in the compactness and simplicity. Thanks to the explicit stiffness matrix computation derived from Bėzier extraction operator, the tedious and complicated full Gaussian quadrature is circumvented, and the storage in elemental stiffness matrices is also significantly simplified for regular design domain occupied with linear elastic material. Similar to the widespread top99 and top88, our code implementation consists of the following parts: a) material properties definition for solid and void material; b) prerequisites for IGA, including the standard solid Bėzier stiffness matrix and element connectivity matrices etc. definitions; c) boundary condition definition; d) fundamental data performing sensitivity filtering; e) optimization parameters initialization; f) topology optimization iteration loop. Moreover, the iteration loop of ITO makes up three main subparts: isogeometric analysis and sensitivity analysis as well as the design variables OC updating process.

According to the numerical examples presented in this article, both iga_top120 and iga_top3D257 are validate to the 2D and 3D TO design problems, respectively. For 2D problem, the converged results by iga_top120 are identical to these obtained from the widespread top88 using sensitivity filter. Subsequently, a comparison is performed between iga_top3D257 and top3D, which reveals the superiorities of iga_top3D257 in both structural performance and optimization efficiency, and advocates the excellent numerical advantages of IGA over FEM in the framework of topology optimization. In the future, the presented code may can be integrated with the artificial neural technique [57] and casted into a very promising approach to performing intelligent design for engineering structures.

# Appendix A

## The 120 lines code for 2D compliance problem.

```matlab
%% A 120 LINE ISOGEOMETRIC TOPOLOGY OPTIMIZATION CODE BASED ON BEZIER EXTRACTION %%
function iga_top120(nelx,nely,volfrac,penal,rmin)
%% MATERIAL PROPERTIES
E0 = 1;
Emin = 1e-3;
nu = 0.3;
%% PREPARE IGA
noPtsX = nelx + 2;
noPtsY = nely + 2;
noCtrPts = noPtsX * noPtsY;
noElems = nelx * nely;
noDofs = 2*noCtrPts;
noPtsX_B = 2*nelx + 1;
noPtsY_B = 2*nely + 1;
noCtrPts_B = noPtsX_B * noPtsY_B;
A11 = [144,-24,-40,24,-36,-28,-8,-20,-12;-24,128,-24,-36,32,-36,-20,0,-20;-40,-24,144,-28,-36,24,-12,-20,-8;
       24,-36,-28,112,-8,-24,24,-36,-28;-36,32,-36,-8,96,-8,-36,32,-36;-28,-36,24,-24,-8,112,-28,-36,24;
       -8,-20,-12,24,-36,-28,144,-24,-40;-20,0,-20,-36,32,-36,-24,128,-24;-12,-20,-8,-28,-36,24,-40,-24,144];
A12 = [45,-30,-15,30,-20,-10,15,-10,-5;30,0,-30,20,0,-20,10,0,-10;15,30,-45,10,20,-30,5,10,-15;
       -30,20,10,0,0,0,30,-20,-10;-20,0,20,0,0,0,0,20,0,-20;-10,-20,30,0,0,0,-10,-20,30;
       -15,10,5,-30,20,10,-45,30,15;-10,0,10,-20,0,20,-30,0,30;-5,-10,15,-10,-20,30,-15,-30,45];
B11 = [-48,-24,-8,24,12,4,24,12,4;-24,-32,-24,12,16,12,12,16,12;-8,-24,-48,4,12,24,4,12,24;
       24,12,4,-48,-24,-8,24,12,4;12,16,12,-24,-32,-24,12,16,12;4,12,24,-8,-24,-48,4,12,24;
       24,12,4,24,12,4,-48,-24,-8;12,16,12,12,16,12,-24,-32,-24;4,12,24,4,12,24,-8,-24,-48];
B12 = [45,90,45,-90,-20,-10,-45,-10,-5;-90,0,90,20,0,-20,10,0,-10;-45,-90,-45,10,20,90,5,10,45;
       90,20,10,0,0,0,-90,-20,-10;-20,0,20,0,0,0,0,20,0,-20;-10,-20,-90,0,0,0,10,20,90;
       45,10,5,90,20,10,-45,-90,-45;-10,0,10,-20,0,20,90,0,-90;-5,-10,-45,-10,-20,-90,45,90,45];
C14 = full(sparse([1 4 7 2 5 8 3 6 9],1:9,ones(1,9)));
KE = E0/(1-nu^2)/360*([A11 A12;A12' C14'*A11*C14]+nu*[B11 B12; B12' C14'*B11*C14]);
Cxi = repmat([0.5,0,0;0.5,1,0.5;0,0,0.5],1,1,nelx);
Cxi(:,1,1) = [1;0;0];
Cxi(:,3,nelx) = [0;0;1];
Cet = repmat([0.5,0,0;0.5,1,0.5;0,0,0.5],1,1,nely);
Cet(:,1,1) = [1;0;0];
Cet(:,3,nely) = [0;0;1];
CxiG = sparse(noPtsX,noPtsX_B);
CxiG(noPtsX*([1 2:2:noPtsX_B-1 noPtsX_B]-1)+(1:noPtsX))=1;
CxiG(noPtsX*([3:2:noPtsX_B-2 3:2:noPtsX_B-2]-1)+[2:noPtsX-2,3:noPtsX-1])=0.5;
CetG = sparse(noPtsY,noPtsY_B);
CetG(noPtsY*([1 2:2:noPtsY_B-1 noPtsY_B]-1)+(1:noPtsY))=1;
CetG(noPtsY*([3:2:noPtsY_B-2 3:2:noPtsY_B-2]-1)+[2:noPtsY-2,3:noPtsY-1])=0.5;
elConnU_B = bsxfun(@plus,(1:2:nelx*2-1)',0:2);
elConnV_B = bsxfun(@plus,(1:2:nely*2-1)',0:2);
element_B = (reshape(permute(repmat(elConnV_B,1,1,nelx,3),[3,1,4,2]),noElems,9)-1)*noPtsX_B+repmat(elConnU_B,nely,3);
edofMat_B = [element_B noCtrPts_B+element_B];
elConnU = bsxfun(@plus,(1:nelx)',0:2);

elConnV = bsxfun(@plus,(1:nely)',0:2);
element = (reshape(permute(repmat(elConnV,1,1,nelx,3),[3,1,4,2]),noElems,9)-1)*noPtsX+repmat(elConnU,nely,3);
edofMat = [element noCtrPts+element];
iK = reshape(kron(edofMat,ones(18,1))',324*noElems,1);
jK = reshape(kron(edofMat,ones(1,18))',324*noElems,1);
clear elConnU_B elConnV_B element_B elConnU elConnV element edofMat
%% BOUNDARY OF HALF MBB BEAM
F = sparse(noCtrPts+noPtsX*(noPtsY-1)+1,1,-1,noDofs,1);
U = zeros(noDofs,1);
fixeddofs = union(1:noPtsX:noCtrPts, noCtrPts+noPtsX);
alldofs = 1:noDofs;
freedofs = setdiff(alldofs,fixeddofs);
%% PREPARE FILTER
iH = ones(noElems*(2*(ceil(rmin)-1)+1)^2,1);
jH = ones(size(iH));
sH = zeros(size(iH));
k = 0;
for i1 = 1:nely
  for j1 = 1:nelx
    e1 = (i1-1)*nelx+j1;
    for i2 = max(i1-(ceil(rmin)-1),1):min(i1+(ceil(rmin)-1),nely)
      for j2 = max(j1-(ceil(rmin)-1),1):min(j1+(ceil(rmin)-1),nelx)
        e2 = (i2-1)*nelx+j2;
        k = k+1;
        iH(k) = e1;
        jH(k) = e2;
        sH(k) = max(0,rmin-sqrt((i1-i2)^2+(j1-j2)^2));
      end
    end
  end
end
H = sparse(iH,jH,sH);
Hs = sum(H,2);
%% INITIALIZE ITERATION
x = repmat(volfrac,nelx,nely);
loop = 0;
change = 1;
while change > 0.01
  loop = loop + 1;
  %% IGA
  sK = zeros(324*noElems,1,1);
  k = 0;
  for j = 1:nely
    for i = 1:nelx
      k = k+1;
      CK = [kron(Cet(:,:,j),Cxi(:,:,i)),zeros(9,9);zeros(9,9),kron(Cet(:,:,j),Cxi(:,:,i))];
```

```matlab
93              sK((k-1)*324+1:k*324) = reshape(CK*KE*CK',324,1)*(Emin+x(i,j).^penal*(E0-Emin));
94          end
95      end
96      K = sparse(iK,jK,sK);
97      K=(K+K')/2;
98      U(freedofs) = K(freedofs,freedofs)\F(freedofs);
99      U_B = full([kron(CetG,CxiG)'*U(1:noCtrPts);kron(CetG,CxiG)'*U(noCtrPts+(1:noCtrPts))]);
100     %% OBJECTIVE FUNCTION CALCULATION AND SENSITIVITY ANALYSIS
101     ce = reshape(sum((U_B(edofMat_B)*KE).*U_B(edofMat_B),2),nelx,nely);
102     c = sum(sum((Emin+x.^penal*(E0-Emin)).*ce));
103     dc = -penal*(E0-Emin)*x.^(penal-1).*ce;
104     dv = ones(nelx,nely);
105     %% FILTERING/MODIFICATION OF SENSITIVITIES
106     dc(:) = H*(x(:).*dc(:))./Hs./max(1e-3,x(:));
107     %% OPTIMALITY CRITERIA UPDATE OF DESIGN VARIABLES AND PHYSICAL DENSITIES
108     l1 = 0; l2 = 1e9; move = 0.2;
109     while (l2-l1)/(l1+l2) > 1e-3
110         lmid = 0.5*(l2+l1);
111         xnew = max(0,max(x-move,min(1,min(x+move,x.*sqrt(-dc./dv/lmid)))));
112         if sum(xnew(:)) > volfrac*noElems, l1 = lmid; else l2 = lmid; end
113     end
114     change = max(abs(xnew(:)-x(:)));
115     x = xnew;
116     %% PRINT RESULTS
117     fprintf(' It.:%5i Obj.:%11.4f Vol.:%7.3f ch.:%7.3f\n',loop,c,mean(x(:)),change);
118     %% PLOT DENSITIES
119     colormap(gray); imagesc(1-flipud(x')); caxis([0 1]); axis equal; axis off; drawnow;
120 end
121 % =================================================================
122 % === This code was written by X XIE and A Yang, School of Advanced    ===
123 % === Manufacturing, Nanchang University, Nanchang, CHINA              ===
124 % === -----------------------------------------------------------      ===
125 % === Please send your suggestions and comments to: xiexd020@ncu.edu.cn ===
126 % === -----------------------------------------------------------      ===
```

## Appendix B

### The 257 lines code for 3D compliance problem.

```matlab
1   %% A 257 LINE 3D ISOGEOMETRIC TOPOLOGY OPTIMIZATION CODE BASED ON BEZIER EXTRACTION %%
2   function iga_top3D257(nelx,nely,nelz,volfrac,penal,rmin)
3   %% MATERIAL PROPERTIES
4   E0 = 1;
5   Emin = 1e-3;
6   nu = 0.3;
7   %% PREPARE IGA
8   noPtsX = nelx + 2;
9   noPtsY = nely + 2;
10  noPtsZ = nelz + 2;
11  noCtrPts   = noPtsX * noPtsY *noPtsZ;
12  noElems = nelx * nely *nelz;
13  noDofs = 3*noCtrPts;
14  noPtsX_B = 2*nelx + 1;
15  noPtsY_B = 2*nely + 1;
16  noPtsZ_B = 2*nelz + 1;
17  noCtrPts_B   = noPtsX_B * noPtsY_B *noPtsZ_B;
18  A11 = [864,144,-48,144,-72,-72,-48,-72,-40,144,-72,-72,-72,-72,-108,-60,-72,-60,-28,-48,-72,-40,-72,-60,-28,-40,-28,-12;
19         144,672,144,-72,144,-72,-72,-16,-72,-72,144,-72,-108,-24,-108,-60,-40,-60,-72,-16,-72,-60,-40,-60,-28,-24,-28;
20         -48,144,864,-72,-72,144,-40,-72,-48,-72,-72,144,-60,-108,-72,-28,-60,-72,-40,-72,-48,-28,-60,-72,-12,-28,-40;
21         144,-72,-72,672,144,-16,144,-72,-72,-72,-108,-60,144,-24,-40,-72,-60,-28,-16,-40,-24,-72,-60,-28;
22         -72,144,-72,144,512,144,-72,144,-72,-108,-24,-108,-24,128,-24,-108,-24,-108,-60,-40,-60,-40,0,-40,-60,-40,-60;
23         -72,-72,144,-16,144,672,-72,-72,144,-60,-108,-72,-40,-24,144,-60,-108,-72,-28,-60,-72,-24,-40,-16,-28,-60,-72;
24         -48,-72,-40,144,-72,-72,864,144,-48,-72,-60,-28,-72,-108,-60,144,-72,-72,-40,-28,-12,-72,-60,-28,-48,-72,-40;
25         -72,-16,-72,-72,144,-72,144,672,144,-60,-40,-60,-108,-24,-108,-72,144,-72,-28,-24,-28,-60,-40,-60,-72,-16,-72;
26         -40,-72,-48,-72,-72,144,-48,144,864,-28,-60,-72,-60,-108,-72,-72,-72,144,-12,-28,-40,-28,-60,-72,-40,-72,-48;
27         144,-72,-72,-72,-108,-60,-72,-60,-28,672,144,-16,144,-24,-40,-16,-40,-24,144,-72,-72,-72,-108,-60,-72,-60,-28;
28         -72,144,-72,-108,-24,-108,-60,-40,-60,144,512,144,-24,128,-24,-40,0,-40,-72,144,-72,-108,-24,-108,-60,-40,-60;
29         -72,-72,144,-60,-108,-72,-28,-60,-72,-16,144,672,-40,-24,144,-24,-40,-16,-72,-72,144,-60,-108,-72,-28,-60,-72;
30         -72,-108,-60,144,-24,-40,-72,-108,-60,144,-24,-40,512,128,0,144,-24,-40,-72,-108,-60,144,-24,-40,-72,-108,-60;
31         -108,-24,-108,-24,128,-24,-108,-24,-108,-24,128,-24,128,384,128,-24,128,-24,-108,-24,-108,-24,128,-24,-108,-24,-108;
32         -60,-108,-72,-40,-24,144,-60,-108,-72,-40,-24,144,0,128,512,-40,-24,144,-60,-108,-72,-40,-24,144,-60,-108,-72;
33         -72,-60,-28,-72,-108,-60,144,-72,-72,-16,-40,-24,144,-24,-40,672,144,-16,-72,-60,-28,-72,-108,-60,144,-72,-72;
34         -60,-40,-60,-108,-24,-108,-72,144,-72,-40,0,-40,-24,128,-24,144,512,144,-60,-40,-60,-108,-24,-108,-72,144,-72;
35         -28,-60,-72,-60,-108,-72,-72,144,-24,-40,-16,-40,-24,144,-16,144,672,-28,-60,-72,-60,-108,-72,-72,-72,144;
36         -48,-72,-40,-72,-60,-28,-40,-28,-12,144,-72,-72,-72,-108,-60,-72,-60,-28,864,144,-48,144,-72,-72,-48,-72,-40;
37         -72,-16,-72,-60,-40,-60,-28,-24,-28,-72,144,-72,-108,-24,-108,-60,-40,-60,144,672,144,-72,144,-72,-72,-16,-72;
38         -40,-72,-48,-28,-60,-72,-12,-28,-40,-72,-72,144,-60,-108,-72,-28,-60,-72,-48,144,864,-72,-72,144,-40,-72,-48;
39         -72,-60,-28,-16,-40,-24,-72,-60,-28,-72,-108,-60,144,-24,-40,-72,-108,-60,144,-72,-72,672,144,-16,144,-72,-72;
40         -60,-40,-60,-40,0,-40,-60,-40,-60,-108,-24,-108,-24,128,-24,-108,-24,-108,-72,144,-72,144,512,144,-72,144,-72;
41         -28,-60,-72,-24,-40,-16,-28,-60,-72,-60,-108,-72,-40,-24,144,-60,-108,-72,-72,-72,144,-16,144,672,-72,-72,144;
42         -40,-28,-12,-72,-60,-28,-48,-72,-40,-72,-60,-28,-72,-108,-60,144,-72,-72,-48,-72,-40,144,-72,-72,864,144,-48;
43         -28,-24,-28,-60,-40,-60,-72,-16,-72,-60,-40,-60,-108,-24,-108,-72,144,-72,-72,-16,-72,-72,144,-72,144,672,144;
44         -12,-28,-40,-28,-60,-72,-40,-72,-48,-28,-60,-72,-72,-108,-72,-72,144,-40,-72,-48,-72,-72,144,-48,144,864];
45  A12 = [270,-180,-90,180,-120,-60,90,-60,-30,135,-90,-45,90,-60,-30,45,-30,-15,45,-30,-15,30,-20,-10,15,-10,-5;
46         180,0,-180,120,0,-120,60,0,-60,90,0,-90,60,0,-60,30,0,-30,30,0,-30,20,0,-20,10,0,-10;
```

```matlab
47              90,180,-270,60,120,-180,30,60,-90,45,90,-135,30,60,-90,15,30,-45,15,30,-45,10,20,-30,5,10,-15;
48              -180,120,60,0,0,0,180,-120,-60,-90,60,30,0,0,0,90,-60,-30,-30,20,10,0,0,0,30,-20,-10;
49              -120,0,120,0,0,0,120,0,-120,-60,0,60,0,0,0,60,0,-60,-20,0,20,0,0,0,20,0,-20;
50              -60,-120,180,0,0,0,60,120,-180,-30,-60,90,0,0,0,30,60,-90,-10,-20,30,0,0,0,10,20,-30;
51              -90,60,30,-180,120,60,-270,180,90,-45,30,15,-90,60,30,-135,90,45,-15,10,5,-30,20,10,-45,30,15;
52              -60,0,60,-120,0,120,-180,0,180,-30,0,30,-60,0,60,-90,0,90,-10,0,10,-20,0,20,-30,0,30;
53              -30,-60,90,-60,-120,180,-90,-180,270,-15,-30,45,-30,-60,90,-45,-90,135,-5,-10,15,-10,-20,30,-15,-30,45;
54              135,-90,-45,90,-60,-30,45,-30,-15,180,-120,-60,120,-80,-40,60,-40,-20,135,-90,-45,90,-60,-30,45,-30,-15;
55              90,0,-90,60,0,-60,30,0,-30,120,0,-120,80,0,-80,40,0,-40,90,0,-90,60,0,-60,30,0,-30;
56              45,90,-135,30,60,-90,15,30,-45,60,120,-180,40,80,-120,20,40,-60,45,90,-135,30,60,-90,15,30,-45;
57              -90,60,30,0,0,0,90,-60,-30,-120,80,40,0,0,0,120,-80,-40,-90,60,30,0,0,0,90,-60,-30;
58              -60,0,60,0,0,0,60,0,-60,-80,0,80,0,0,0,80,0,-80,-60,0,60,0,0,0,60,0,-60;
59              -30,-60,90,0,0,0,30,60,-90,-40,-80,120,0,0,0,40,80,-120,-30,-60,90,0,0,0,30,60,-90;
60              -45,30,15,-90,60,30,-135,90,45,-60,40,20,-120,80,40,-180,120,60,-45,30,15,-90,60,30,-135,90,45;
61              -30,0,30,-60,0,60,-90,0,90,-40,0,40,-80,0,80,-120,0,120,-30,0,30,-60,0,60,-90,0,90;
62              -15,-30,45,-30,-60,90,-45,-90,135,-20,-40,60,-40,-80,120,-60,-120,180,-15,-30,45,-30,-60,90,-45,-90,135;
63              45,-30,-15,30,-20,-10,15,-10,-5,135,-90,-45,90,-60,-30,45,-30,-15,270,-180,-90,180,-120,-60,90,-60,-30;
64              30,0,-30,20,0,-20,10,0,-10,90,0,-90,60,0,-60,30,0,-30,180,0,-180,120,0,-120,60,0,-60;
65              15,30,-45,10,20,-30,5,10,-15,45,90,-135,30,60,-90,15,30,-45,90,180,-270,60,120,-180,30,60,-90;
66              -30,20,10,0,0,0,30,-20,-10,-90,60,30,0,0,0,90,-60,-30,-180,120,60,0,0,0,180,-120,-60;
67              -20,0,20,0,0,0,20,0,-20,-60,0,60,0,0,0,60,0,-60,-120,0,120,0,0,0,120,0,-120;
68              -10,-20,30,0,0,0,10,20,-30,-30,-60,90,0,0,0,30,60,-90,-60,-120,180,0,0,0,60,120,-180;
69              -15,10,5,-30,20,10,-45,30,15,-45,30,15,-90,60,30,-135,90,45,-90,60,30,-180,120,60,-270,180,90;
70              -10,0,10,-20,0,20,-30,0,30,-30,0,30,-60,0,60,-90,0,90,-60,0,60,-120,0,120,-180,0,180;
71              -5,-10,15,-10,-20,30,-15,-30,45,-15,-30,45,-30,-60,90,-45,-90,135,-30,-60,90,-60,-120,180,-90,-180,270];
72     B11 = [-1440,-432,-48,-144,72,72,144,120,56,-144,72,72,216,180,84,168,108,44,144,120,56,168,108,44,88,52,20;
73              -432,-1056,-432,72,-144,72,120,80,120,72,-144,72,180,120,180,108,104,108,120,80,120,108,104,108,52,56,52;
74              -48,-432,-1440,72,72,-144,56,120,144,72,72,-144,84,180,216,44,108,168,56,120,144,44,108,168,20,52,88;
75              -144,72,72,-1152,-384,-64,-144,72,72,216,180,84,-192,0,32,216,180,84,168,108,44,64,64,32,168,108,44;
76              72,-144,72,-384,-832,-384,72,-144,72,180,120,180,0,-160,0,180,120,180,108,104,108,64,32,64,108,104,108;
77              72,72,-144,-64,-384,-1152,72,72,-144,84,180,216,32,0,-192,84,180,216,44,108,168,32,64,64,44,108,168;
78              144,120,56,-144,72,72,-1440,-432,-48,168,108,44,216,180,84,-144,72,72,88,52,20,168,108,44,144,120,56;
79              120,80,120,72,-144,72,-432,-1056,-432,108,104,108,180,120,180,72,-144,72,52,56,52,108,104,108,120,80,120;
80              56,120,144,72,72,-144,-48,-432,-1440,44,108,168,84,180,216,72,72,-144,20,52,88,44,108,168,56,120,144;
81              -144,72,72,216,180,84,168,108,44,-1152,-384,-64,-192,0,32,64,64,32,-144,72,72,216,180,84,168,108,44;
82              72,-144,72,180,120,180,108,104,108,-384,-832,-384,0,-160,0,64,32,64,72,-144,72,180,120,180,108,104,108;
83              72,72,-144,84,180,216,44,108,168,-64,-384,-1152,32,0,-192,32,64,64,72,72,-144,84,180,216,44,108,168;
84              216,180,84,-192,0,32,216,180,84,-192,0,32,-896,-320,-64,-192,0,32,216,180,84,-192,0,32,216,180,84;
85              180,120,180,0,-160,0,180,120,180,0,-160,0,-320,-640,-320,0,-160,0,180,120,180,0,-160,0,180,120,180;
86              84,180,216,32,0,-192,84,180,216,32,0,-192,-64,-320,-896,32,0,-192,84,180,216,32,0,-192,84,180,216;
87              168,108,44,216,180,84,-144,72,72,64,64,32,-192,0,32,-1152,-384,-64,168,108,44,216,180,84,-144,72,72;
88              108,104,108,180,120,180,72,-144,72,64,32,64,0,-160,0,-384,-832,-384,108,104,108,180,120,180,72,-144,72;
89              44,108,168,84,180,216,72,72,-144,32,64,64,32,0,-192,-64,-384,-1152,44,108,168,84,180,216,72,72,-144;
90              144,120,56,168,108,44,88,52,20,-144,72,72,216,180,84,168,108,44,-1440,-432,-48,-144,72,72,144,120,56;
91              120,80,120,108,104,108,52,56,52,72,-144,72,180,120,180,108,104,108,-432,-1056,-432,72,-144,72,120,80,120;
92              56,120,144,44,108,168,20,52,88,72,72,-144,84,180,216,44,108,168,-48,-432,-1440,72,72,-144,56,120,144;
93              168,108,44,64,64,32,168,108,44,216,180,84,-192,0,32,216,180,84,-144,72,72,-1152,-384,-64,-144,72,72;
94              108,104,108,64,32,64,108,104,108,180,120,180,0,-160,0,180,120,180,72,-144,72,-384,-832,-384,72,-144,72;
95              44,108,168,32,64,64,44,108,168,84,180,216,32,0,-192,84,180,216,72,72,-144,-64,-384,-1152,72,72,-144;
96              88,52,20,168,108,44,144,120,56,168,108,44,216,180,84,-144,72,72,144,120,56,-144,72,72,-1440,-432,-48;
97              52,56,52,108,104,108,120,80,120,108,104,108,180,120,180,72,-144,72,120,80,120,72,-144,72,-432,-1056,-432;
98              20,52,88,44,108,168,56,120,144,44,108,168,84,180,216,72,72,-144,56,120,144,72,72,-144,-48,-432,-1440];
99     B12 = [-270,540,270,-540,120,60,-270,60,30,-135,270,135,-270,60,30,-135,30,15,-45,90,45,-90,20,10,-45,10,5;
100             -540,0,540,-120,0,120,-60,0,60,-270,0,270,-60,0,60,-30,0,30,-90,0,90,-20,0,20,-10,0,10;
101             -270,-540,270,-60,-120,540,-30,-60,270,-135,-270,135,-30,-60,270,-15,-30,135,-45,-90,45,-10,-20,90,-5,-10,45;
102             540,-120,-60,0,0,0,-540,120,60,270,-60,-30,0,0,0,-270,60,30,90,-20,-10,0,0,0,-90,20,10;
103             120,0,-120,0,0,0,-120,0,120,60,0,-60,0,0,0,-60,0,60,20,0,-20,0,0,0,-20,0,20;
104             60,120,-540,0,0,0,-60,-120,540,30,60,-270,0,0,0,-30,-60,270,10,20,-90,0,0,0,-10,-20,90;
105             270,-60,-30,540,-120,-60,270,-540,-270,135,-30,-15,270,-60,-30,135,-270,-135,45,-10,-5,90,-20,-10,45,-90,-45;
106             60,0,-60,120,0,-120,540,0,-540,30,0,-30,60,0,-60,270,0,-270,10,0,-10,20,0,-20,90,0,-90;
107             30,60,-270,60,120,-540,270,540,-270,15,30,-135,30,60,-270,135,270,-135,5,10,-45,10,20,-90,45,90,-45;
108             -135,270,135,-270,60,30,-135,30,15,-180,360,180,-360,80,40,-180,40,20,-135,270,135,-270,60,30,-135,30,15;
109             -270,0,270,-60,0,60,-30,0,30,-360,0,360,-80,0,80,-40,0,40,-270,0,270,-60,0,60,-30,0,30;
110             -135,-270,135,-30,-60,270,-15,-30,135,-180,-360,180,-40,-80,360,-20,-40,180,-135,-270,135,-30,-60,270,-15,-30,135;
111             270,-60,-30,0,0,0,-270,60,30,360,-80,-40,0,0,0,-360,80,40,270,-60,-30,0,0,0,-270,60,30;
112             60,0,-60,0,0,0,-60,0,60,80,0,-80,0,0,0,-80,0,80,60,0,-60,0,0,0,-60,0,60;
113             30,60,-270,0,0,0,-30,-60,270,40,80,-360,0,0,0,-40,-80,360,30,60,-270,0,0,0,-30,-60,270;
114             135,-30,-15,270,-60,-30,135,-270,-135,180,-40,-20,360,-80,-40,180,-360,-180,135,-30,-15,270,-60,-30,135,-270,-135;
115             30,0,-30,60,0,-60,270,0,-270,40,0,-40,80,0,-80,360,0,-360,30,0,-30,60,0,-60,270,0,-270;
116             15,30,-135,30,60,-270,135,270,-135,20,40,-180,40,80,-360,180,360,-180,15,30,-135,30,60,-270,135,270,-135;
117             -45,90,45,-90,20,10,-45,10,5,-135,270,135,-270,60,30,-135,30,15,-270,540,270,-540,120,60,-270,60,30;
118             -90,0,90,-20,0,20,-10,0,10,-270,0,270,-60,0,60,-30,0,30,-540,0,540,-120,0,120,-60,0,60;
119             -45,-90,45,-10,-20,90,-5,-10,45,-135,-270,135,-30,-60,270,-15,-30,135,-270,-540,270,-60,-120,540,-30,-60,270;
120             90,-20,-10,0,0,0,-90,20,10,270,-60,-30,0,0,0,-270,60,30,540,-120,-60,0,0,0,-540,120,60;
121             20,0,-20,0,0,0,-20,0,20,60,0,-60,0,0,0,-60,0,60,120,0,-120,0,0,0,-120,0,120;
122             10,20,-90,0,0,0,-10,-20,90,30,60,-270,0,0,0,-30,-60,270,60,120,-540,0,0,0,-60,-120,540;
123             45,-10,-5,90,-20,-10,45,-90,-45,135,-30,-15,270,-60,-30,135,-270,-135,270,-60,-30,540,-120,-60,270,-540,-270;
124             10,0,-10,20,0,-20,90,0,-90,30,0,-30,60,0,-60,270,0,-270,60,0,-60,120,0,-120,540,0,-540;
125             5,10,-45,10,20,-90,45,90,-45,15,30,-135,30,60,-270,135,270,-135,30,60,-270,60,120,-540,270,540,-270];
126     C1 = full(sparse(1:27,[1,2,3,10,11,12,19,20,21,4,5,6,13,14,15,22,23,24,7,8,9,16,17,18,25,26,27],ones(1,27)));
127     C2 = full(sparse(1:27,[1,4,7,10,13,16,19,22,25,2,5,8,11,14,17,20,23,26,3,6,9,12,15,18,21,24,27],ones(1,27)));
128     C3 = full(sparse(1:27,[1,10,19,4,13,22,7,16,25,2,11,20,5,14,23,8,17,26,3,12,21,6,15,24,9,18,27],ones(1,27)));
129     KE = E0/(1+nu)/(1-2*nu)/5400*([A11,A12,C1'*A12*C1;A12',A11,C2'*A12*C2;C1'*A12'*C1,C2'*A12'*C2,A11]+...
130          nu*[B11,B12,C1'*B12*C1;B12',C2'*B11*C2,C2'*B12*C2;C1'*B12'*C1,C2'*B12'*C2,C3'*B11*C3]);
131     Cxi = repmat([0.5,0,0;0.5,1,0.5;0,0,0.5],1,1,nelx);
132     Cxi(:,1,1) = [1;0;0];
133     Cxi(:,3,nelx) = [0;0;1];
134     Cet = repmat([0.5,0,0;0.5,1,0.5;0,0,0.5],1,1,nely);
135     Cet(:,1,1) = [1;0;0];
136     Cet(:,3,nely) = [0;0;1];
137     Cze = repmat([0.5,0,0;0.5,1,0.5;0,0,0.5],1,1,nelz);
138     Cze(:,1,1) = [1;0;0];
```

```matlab
    Cze(:,3,nelz) = [0;0;1];
    CxiG = sparse(noPtsX,noPtsX_B);
    CxiG(noPtsX*([1 2:2:noPtsX_B-1 noPtsX_B]-1)+(1:noPtsX))=1;
    CxiG(noPtsX*([3:2:noPtsX_B-2 3:2:noPtsX_B-2]-1)+[2:noPtsX-2,3:noPtsX-1])=0.5;
    CetG = sparse(noPtsY,noPtsY_B);
    CetG(noPtsY*([1 2:2:noPtsY_B-1 noPtsY_B]-1)+(1:noPtsY))=1;
    CetG(noPtsY*([3:2:noPtsY_B-2 3:2:noPtsY_B-2]-1)+[2:noPtsY-2,3:noPtsY-1])=0.5;
    CzeG = sparse(noPtsZ,noPtsZ_B);
    CzeG(noPtsZ*([1 2:2:noPtsZ_B-1 noPtsZ_B]-1)+(1:noPtsZ))=1;
    CzeG(noPtsZ*([3:2:noPtsZ_B-2 3:2:noPtsZ_B-2]-1)+[2:noPtsZ-2,3:noPtsZ-1])=0.5;
    elConnU_B = bsxfun(@plus,(1:2:nelx*2-1)',0:2);
    elConnV_B = bsxfun(@plus,(1:2:nely*2-1)',0:2);
    elConnW_B = bsxfun(@plus,(1:2:nelz*2-1)',0:2);
    element_B = (reshape(permute(repmat(elConnW_B,1,1,nelx,nely,3,3),[3 4 1 5 6 2]),noElems,27)-1)*noPtsY_B*noPtsX_B+...
        (reshape(permute(repmat(elConnV_B,1,1,nelx,nelz,3,3),[3 1 4 5 2 6]),noElems,27)-1)*noPtsX_B+...
        reshape(permute(repmat(elConnU_B,1,1,nely,nelz,3,3),[1 3 4 2 5 6]),noElems,27);
    edofMat_B = [element_B noCtrPts_B+element_B 2*noCtrPts_B+element_B];
    elConnU = bsxfun(@plus,(1:nelx)',0:2);
    elConnV = bsxfun(@plus,(1:nely)',0:2);
    elConnW = bsxfun(@plus,(1:nelz)',0:2);
    element = (reshape(permute(repmat(elConnW,1,1,nelx,nely,3,3),[3 4 1 5 6 2]),noElems,27)-1)*noPtsY*noPtsX+...
        (reshape(permute(repmat(elConnV,1,1,nelx,nelz,3,3),[3 1 4 5 2 6]),noElems,27)-1)*noPtsX+...
        reshape(permute(repmat(elConnU,1,1,nely,nelz,3,3),[1 3 4 2 5 6]),noElems,27);
    edofMat = [element noCtrPts+element 2*noCtrPts+element];
    iK = reshape(kron(edofMat,ones(81,1))',6561*noElems,1);
    jK = reshape(kron(edofMat,ones(1,81))',6561*noElems,1);
    clear elConnU_B elConnV_B elConnW_B element_B elConnU elConnV elConnW element edofMat
%% BOUNDARY OF CANTILEVER BEAM
    ind = [noPtsX*floor(noPtsY/2)+noPtsX*noPtsY*floor(noPtsZ/2) noPtsX*floor(noPtsY/2)+noPtsX*noPtsY*(floor(noPtsZ/2)+1)...
        noPtsX*(floor(noPtsY/2)+1)+noPtsX*noPtsY*floor(noPtsZ/2)  noPtsX*(floor(noPtsY/2)+1)+noPtsX*noPtsY*(floor(noPtsZ/2)+1)];
    F = sparse(ind+2*noCtrPts,1,-1/numel(ind),noDofs,1);
    U = zeros(noDofs,1);
    fixeddofs = 1:noPtsX:noDofs;
    alldofs = 1:noDofs;
    freedofs = setdiff(alldofs,fixeddofs);
%% PREPARE FILTER
    iH = ones(noElems*(2*(ceil(rmin)-1)+1)^3,1);
    jH = ones(size(iH));
    sH = zeros(size(iH));
    k = 0;
    for i1 = 1:nelz
        for j1 = 1:nely
            for k1 = 1:nelx
                e1 = (i1-1)*nely*nelx+(j1-1)*nelx+k1;
                for i2 = max(i1-(ceil(rmin)-1),1):min(i1+(ceil(rmin)-1),nelz)
                    for j2 = max(j1-(ceil(rmin)-1),1):min(j1+(ceil(rmin)-1),nely)
                        for k2 = max(k1-(ceil(rmin)-1),1):min(k1+(ceil(rmin)-1),nelx)
                            e2 = (i2-1)*nely*nelx+(j2-1)*nelx+k2;
                            k = k+1;
                            iH(k) = e1;
                            jH(k) = e2;
                            sH(k) = max(0,rmin-sqrt((i1-i2)^2+(j1-j2)^2+(k1-k2)^2));
                        end
                    end
                end
            end
        end
    end
    H = sparse(iH,jH,sH);
    Hs = sum(H,2);
%% INITIALIZE ITERATION
    x = repmat(volfrac,nelx,nely,nelz);
    loop = 0;
    change = 1;
    while change > 0.01
        loop = loop + 1;
%% IGA
        sK = zeros(6561*noElems,1,1);
        m = 0;
        for k = 1:nelz
            for j = 1:nely
                for i = 1:nelx
                    m = m+1;
                    CK = [kron(Cze(:,:,k),kron(Cet(:,:,j),Cxi(:,:,i))),zeros(27,2*27);...
                        zeros(27,27),kron(Cze(:,:,k),kron(Cet(:,:,j),Cxi(:,:,i))),zeros(27,27);...
                        zeros(27,2*27),kron(Cze(:,:,k),kron(Cet(:,:,j),Cxi(:,:,i)))];
                    sK((m-1)*6561+1:m*6561) = reshape(CK*KE*CK',6561,1)*(Emin+x(i,j,k).^penal*(E0-Emin));
                end
            end
        end
        K = sparse(iK,jK,sK);
        K = (K+K')/2;
        U(freedofs) = K(freedofs,freedofs)\F(freedofs);
        U_B = full([kron(CzeG,kron(CetG,CxiG))'*U(1:noCtrPts);kron(CzeG,kron(CetG,CxiG))'*U(noCtrPts+(1:noCtrPts));...
            kron(CzeG,kron(CetG,CxiG))'*U(2*noCtrPts+(1:noCtrPts))]);
%% OBJECTIVE FUNCTION AND SENSITIVITY ANALYSIS
        ce = reshape(sum((U_B(edofMat_B)*KE).*U_B(edofMat_B),2),nelx,nely,nelz);
        c = sum(sum(sum((Emin+x.^penal*(E0-Emin)).*ce)));
        dc = -penal*(E0-Emin)*x.^(penal-1).*ce;
        dv = ones(nelx,nely,nelz);
%% FILTERING/MODIFICATION OF SENSITIVITIES
        dc(:) = H*(x(:).*dc(:))./Hs./max(1e-3,x(:));
```

```
231         %% OPTIMALITY CRITERIA UPDATE OF DESIGN VARIABLES AND PHYSICAL DENSITIES
232         l1 = 0; l2 = 1e9; move = 0.2;
233         while (l2-l1)/(l1+l2) > 1e-3
234             lmid = 0.5*(l2+l1);
235             xnew = max(0,max(x-move,min(1,min(x+move,x.*sqrt(-dc./dv/lmid)))));
236             if sum(xnew(:)) > volfrac*noElems, l1 = lmid; else l2 = lmid; end
237         end
238         change = max(abs(xnew(:)-x(:)));
239         x = xnew;
240         %% PRINT RESULTS
241         fprintf(' It.:%5i Obj.:%11.4f Vol.:%7.3f ch.:%7.3f\n',loop,c,mean(x(:)),change);
242     end
243     %% PLOT RESULT
244     face = [1 2 3 4; 2 6 7 3; 4 3 7 8; 1 5 8 4; 1 2 6 5; 5 6 7 8];
245     set(gcf,'Name','ISO display','NumberTitle','off');
246     for k = 1:nelz
247         for j = 1:nely
248             for i = 1:nelx
249                 if (x(i,j,k) > 0.5)  % User-defined display density threshold
250                     vert = [i-1 j-1 k-1; i-1 j k-1; i j k-1; i j-1 k-1; i-1 j-1 k;i-1 j k; i j k;i j-1 k];
251                     patch('Faces',face,'Vertices',vert,'FaceColor',[0.2+0.8*(1-x(i,j,k)),0.2+0.8*(1-x(i,j,k)),0.2+0.8*(1-x(i,j,k))]);
252                     hold on;
253                 end
254             end
255         end
256     end
257     axis equal; axis tight; axis off; box on; view([30,30]); pause(1e-6);
258     % ==========================================================
259     % === This code was written by X XIE and A Yang, School of Advanced    ===
260     % === Manufacturing, Nanchang University, Nanchang, CHINA              ===
261     % === ----------------------------------------------------------       ===
262     % === Please send your suggestions and comments to: xiexd020@ncu.edu.cn ===
263     % === ----------------------------------------------------------       ===
```

## Declarations

### Ethics approval and consent to participate

Not applicable.

### Consent for publication

Not applicable.

### Funding

This study was supported by funding from the National Natural Science Foundation of China (No. 52205267), and the National Key R&D Program of China (No. 2023YFB2504601).

### Availability of data and materials

The data that support the findings of this study are available from the author, upon reasonable request.

### Competing interests

The authors declare that they have no competing interests.

### Authors' contributions

Xianda Xie wrote the original draft and the provided the methodology. Zhihui Ou performed the data curation. Shuting Writing reviewed the draft and provided the financial support. Qingtian Xie finished the code implementation of iga_top120. Xiaobing Li performed the figures generation. Aodi Yang implemented the iga_top3D257 code. All authors read and approved the final manuscript.

### Acknowledgments

Not applicable.